\documentclass{iopart}
\usepackage{graphicx}
\usepackage{amssymb}
\begin{document}
\title[Relativistic hydrodynamics]{Relativistic hydrodynamics for heavy-ion collisions}  
\author{Jean-Yves Ollitrault}
\address{Service de Physique Th\'eorique, CEA/DSM/SPhT,
  CNRS/MPPU/URA2306\\ CEA Saclay, F-91191 Gif-sur-Yvette Cedex.}
\ead{jean-yves.ollitrault@cea.fr}

\date{\today}
\begin{abstract}
Relativistic hydrodynamics is essential to our current understanding
of nucleus-nucleus collisions at ultrarelativistic energies (current
experiments at the Relativistic Heavy Ion Collider, forthcoming
experiments at the CERN Large Hadron Collider). This is an
introduction to relativistic hydrodynamics for graduate students. 
It includes a detailed derivation of the equations, and a description 
of the hydrodynamical evolution of a heavy-ion collisions. 
Some knowledge of thermodynamics and special
relativity is assumed.  
\end{abstract} 
\section{Introduction}

The use of relativistic hydrodynamics in the context of high-energy
physics dates back to Landau~\cite{Landau:1953gs}, long before QCD was
discovered. High-energy collisions produce  many hadrons of different
sorts going into all directions. One expected that tools from
statistical physics would 
shed light on this complexity. The light eventually came from the
deep-inelastic scattering of electrons, which led to the parton model,
and hydrodynamics was of little use. 
It is only in recent years, with the advent of heavy-ion experiments
at RHIC, that the interest in relativistic hydrodynamics has been
revived. One of the first RHIC papers~\cite{Ackermann:2000tr} 
reported on the ``observation of a higher degree of thermalization
than at lower collision energies''. Several phenomena were observed
which suggested that the matter produced in these collisions behaves
collectively, like a fluid. It was even claimed in 2005 that the RHIC
experiments had created a ``perfect liquid'', with the lowest possible
viscosity. 

Relativistic hydrodynamics is interesting because it is simple and 
general. It is simple because the information on the system is encoded
in its thermodynamic properties, i.e., its equation of state. 
Hydrodynamics is also general, in the sense that it relies on only one
assumption, unfortunately a very strong one: local thermodynamic
equilibrium. No other assumption is made concerning the nature of the
particles and fields, their interactions, the classical/quantum nature
of the phenomena involved. 

This paper is an introduction to relativistic
hydrodynamics in relation with heavy-ion collisions. 
Relativistic hydrodynamics {\it per se\/} is textbook material
since Landau's course has appeared (see Chapter XV of \cite{Landau}).
As for its applications to heavy-ion collisions, they are covered by 
several recent
reviews~\cite{Huovinen:2006jp,Huovinen:2003fa,Kolb:2003dz}. 
The recent experimental achievements show us the old textbook results
under a new perspective, since it appears possible to produce a
relativistic fluid in the laboratory. My purpose is to provide
graduate students with an elementary, self-contained survey of this
exciting research field. 
Sec.~\ref{s:thermo} recalls basic results of thermodynamics and
statistical physics which are commonly used in the context of
hydrodynamics. 
Sec.~\ref{s:hydro} derives the equations of inviscid hydrodynamics. 
Sec.~\ref{s:heavyions} describes the hydrodynamical
evolution of a heavy-ion collision;  the details may be skipped upon first
reading. Sec.~\ref{s:spectra} derives some 
observables which are used as signatures of hydrodynamical
behaviour. 
Finally, Sec.~\ref{s:last} briefly describes the domain of validity of
hydrodynamics. The readers who are interested in relativistic
hydrodynamics in itself should try and work out the problems 
given in appendix.

\section{Thermodynamics}
\label{s:thermo}

We first recall standard identities of thermodynamics and statistical
physics, which are often used in hydrodynamical models. 

\subsection{General identities}

The differential of internal energy is given by the thermodynamic identity
\begin{equation}
dU=-PdV+TdS+\mu dN,
\label{dE}
\end{equation}
where $P$ is the pressure, $V$ the volume, 
$S$ is the entropy, $T$ the temperature, $\mu$ the chemical potential. 
In nonrelativistic systems, $N$ is generally the number of particles,
which is conserved.  
In a relativistic 
system, the number of particles is not conserved: it is always possible
to create a particle-antiparticle pair, provided energy is available. In this case, $N$ no
longer denotes a number of particles, but a conserved quantity, such as the baryon number. 
If there are several conserved quantities $N_i$, one need simply replace $\mu dN$ with 
$\sum_i\mu_i dN_i$. In these notes, we refer to $N$ as to the baryon
number, and to $\mu$ as the baryon chemical potential, but $N$ can be
any conserved quantity. 
The second important difference in relativistic systems is that the 
mass energy $mc^2$ is included in the internal energy. 

The first two terms in the right-hand side of (\ref{dE}) have transparent 
physical interpretations as the elementary work and heat transferred to the system, 
respectively. The third term is mathematically as simple as the two
first terms, but 
lacks such a simple interpretation \footnote{It does in fact have a simple interpretation 
in the more complex situation where a particle is exchanged between
two different systems, e.g., between two solutions with different
concentrations at the same pressure and  
temperature, as occurs in chemistry. It then plays the role of a thermodynamic potential, 
in the sense that it chooses the lowest possible value.}.

The energy is an extensive function of the extensive variables $V$, $S$, $N$, which means that
\begin{equation}
U(\lambda V,\lambda S,\lambda N)=\lambda U(V,S,N).
\label{Extensive}
\end{equation}
Differentiating with respect to $\lambda$, taking $\lambda=1$, 
and using (\ref{dE}), one obtains
\begin{equation}
U=-PV+TS+\mu N.
\label{Extensive2}
\end{equation}
Differentiating this equation and using again (\ref{dE}), 
one obtains the Gibbs-Duhem relation
\begin{equation}
VdP=SdT+Nd\mu
\label{GibbsDuhem}
\end{equation}

In hydrodynamics, the useful quantities 
are not the total energy, entropy and baryon number, but 
rather their densities per unit volume, the energy density $\epsilon\equiv U/V$, the entropy 
density $s\equiv S/V$, and the baryon density $n\equiv N/V$. All these densities are intensive 
quantities. (\ref{Extensive2}) and (\ref{GibbsDuhem}) give respectively
\begin{equation}
\epsilon=-P+Ts+\mu n.
\label{Extensive3}
\end{equation}
and 
\begin{equation}
dP=sdT+nd\mu.
\label{dP}
\end{equation}
Differentiating (\ref{Extensive3}) and using (\ref{dP}), one obtains
\begin{equation}
d\epsilon=Tds+\mu dn.
\label{depsilon}
\end{equation} 
These identities will be used extensively below.

\subsection{Baryonless fluid}

If the baryon density $n$ vanishes throughout the fluid, the 
corresponding terms disappear from
(\ref{Extensive3}-\ref{depsilon}). The same holds if the chemical
potential $\mu$ is zero throughout the fluid. This shows that ``zero
baryon density'' is in practice equivalent to ``no conserved baryon
number''. Such a fluid has only one intensive degree of freedom.  

The fluid produced in a heavy-ion collision has three conserved
charges, which are the net number of quarks (i.e., number of quarks
minus number of antiquarks) of each flavour $u$, $d$, $s$. 
There is an excess of $u$ and $d$ quarks over antiquarks because of
the incoming nuclei. However, this excess turns out to be negligible at
ultrarelativistic energies because the number of produced particles
overwhelms the number of incoming nucleons. In practice, doing
a hydro calculation with $n=0$ is a rough
approximation, but a reasonable one. 

\subsection{Isentropic process}

The entropy of an inviscid fluid is conserved throughout its
evolution, as we shall see in Sec.~\ref{s:hydro}. This is why 
isentropic processes are important. In an isentropic process, 
both $S$ and $N$ are conserved, and only the volume $V$ changes. The
variations of entropy density and baryon density are given by  
\begin{equation}
\frac{ds}{s}=\frac{dn}{n}=-\frac{dV}{V}
\end{equation}
To compute the variation of energy density, we use  
(\ref{dE}), which reduces to $dU =-PdV$:
\begin{equation}
dU=d(\epsilon V)=\epsilon dV+Vd\epsilon =-PdV,
\end{equation}
hence
\begin{equation}
\frac{d\epsilon }{\epsilon +P}=-\frac{dV}{V}=\frac{ds}{s}=\frac{dn}{n}.
\label{isentropic}
\end{equation}

\subsection{Classical ideal gas}
\label{s:kinetictheory}

An ideal gas is made of independent particles, and is best described by the 
grand-canonical ensemble of statistical mechanics. We choose the natural system
of units $k_B=1$ (one recovers the conventional unit system by
replacing everywhere $T\to k_B T$ and $S\to S/k_B$ in the expressions below).
For simplicity, we consider a gas made of identical, spinless
particles, each of which carries a baryon number equal to unity (although
such particles do not exist). 

In a finite volume $V$, the values of the momentum $\vec p$ are discrete 
(from quantum mechanics). 
The average number of particles with momentum 
$\vec p$ is $1/(\exp((E_{\vec p}-\mu)/T)\pm 1)$, where $E_{\vec
  p}\equiv \sqrt{\vec p^2+m^2}$ is the particle energy (we choose the
natural unit system where $c=1$), 
and the sign depends on whether the particle is a fermion or a boson. For 
sake of simplicity, we take the classical limit where this number is much 
smaller than unity: both Bose-Einstein and Fermi-Dirac 
statistics then reduce to Maxwell-Boltzmann statistics:
\begin{equation}
\frac{1}{e^{(E_{\vec p}-\mu)/T}\pm 1}\simeq e^{(-E_{\vec p}+\mu)/T} \ll 1.
\label{maxwell}
\end{equation}
The particle density, energy density and kinetic pressure are random  
variables in the grand-canonical ensemble. 
Their average values are 
\begin{eqnarray}
n&=&\frac{1}{V}\sum_{\vec p}e^{(-E_{\vec p}+\mu)/T} \cr
\epsilon &=&\frac{1}{V}\sum_{\vec p}E_{\vec p}\, e^{(-E_{\vec p}+\mu)/T} \cr
P&=&\frac{1}{V}\sum_{\vec p}p_x v_x e^{(-E_{\vec p}+\mu)/T},
\label{kineticth}
\end{eqnarray}
where $p_x$ and $v_x$ denote the components of the particle momentum
and velocity along an arbitrary axis $x$. 
This expression of the kinetic pressure is obtained by evaluating the total 
momentum transferred per unit time by elastic collisions with a unit surface  
perpendicular to the $x$-axis. In other terms, it is the momentum flux
along $x$.  
This definition will be used later.  
For a large volume, the sum over momenta is written as an integral:
\begin{equation}
\frac{1}{V}\sum_{\vec p} \to \int \frac{d^3 p}{(2\pi \hbar)^3}.
\label{riemannsum}
\end{equation}

It is an instructive exercise to check that the kinetic pressure coincides
with the thermodynamic pressure, i.e., that it satisfies the expected 
thermodynamic identities. The Gibbs-Duhem relation, (\ref{dP}), gives
$n=(\partial P/\partial\mu)_T$. On the other hand, the kinetic pressure
(\ref{kineticth}) satisfies $(\partial P/\partial\mu)_T=P/T$. 
Putting together these two relations, we obtain 
\begin{equation}
P=n T,
\label{clapeyron}
\end{equation}
which is nothing but the ideal gas law. However, it is not obvious that $P$ and 
$n$ defined by (\ref{kineticth}) satisfy (\ref{clapeyron}). This requires
a little algebra.  The velocity $v_x$ 
is given by Hamilton's 
equation $v_x=\partial E_{\vec p}/\partial p_x$. One then writes
\begin{equation}
v_x e^{(-E_{\vec p}+\mu)/T}=-T \frac{\partial}{\partial p_x}\left(e^{(-E_{\vec p}+\mu)/T}\right).
\label{partintegral}
\end{equation}
Inserting this identity into (\ref{kineticth}), and integrating 
by parts over the variable $p_x$, one recovers (\ref{clapeyron}).

In order to compute the pressure, one uses rotational symmetry of the
integrand in (\ref{kineticth}), and one replaces $p_xv_x$ 
with $\vec p\cdot\vec v/3=pv/3$. For massless particles, this 
gives immediately
\begin{equation}
P=\frac{\epsilon }{3}.
\label{blackbody}
\end{equation}
This relation holds approximately for a quark-gluon plasma at high
temperatures, where interactions are small due to asymptotic freedom. 

The integrals in (\ref{kineticth}) can easily be evaluated for
massless particles.
For a baryonless quark-gluon plasma ($\mu=0$), 
this gives 
\begin{eqnarray}
n&=&\frac{g}{\pi^2\hbar^3} T^3\cr
\epsilon&=& 3 P=3 n T, 
\label{idealqgp}
\end{eqnarray}
where $g$ is the number of degrees of freedom (spin+colour+flavour), 
16 for gluons and 24 for light $u$ and $d$ quarks, i.e., $g\approx
40$. 
Note that $n$ denotes here the particle density, not the baryon
density. (\ref{Extensive3}) gives $\epsilon+P=T s$, so that 
\begin{equation}
s=4 n.
\label{entropyperparticle}
\end{equation}
The entropy per particle is approximately 4 in a quark-gluon plasma. 
(the ratio is in fact 3.6 for bosons, 4.2 for fermions.)

For nonrelativistic particles, note that 
$P\ll \epsilon $. This is because $\epsilon $ includes the 
huge mass energy $mc^2$.

\section{Equations of relativistic hydrodynamics}
\label{s:hydro}

Standard thermodynamics is about a system in global thermodynamic equilibrium. 
This means that intensive parameters ($P$, $T$, $\mu$) are constant 
throughout the volume, and also that the system is globally at rest, 
which means that its total momentum is 0. In this section, we study 
systems whose pressure and temperature vary with space and time, and
which are not at rest, such as the indian atmosphere during monsoon. 
We however request that the system is in {\it local\/} thermodynamic equilibrium,
which means that pressure and temperature are varying so slowly that for any point, 
one can assume thermodynamic equilibrium in some neighbourhood 
about that point.
Here, ``neighbourhood'' has the same meaning as in mathematics, and there is no 
prescription as to the actual size of this neighbourhood, or ``fluid element''. 
There is, however, a general condition for local thermodynamic equilibrium to apply,
which is that the  mean free path of a particle between two collisions is much smaller
than all the characteristic dimensions of the system. We come back to this important
issue in Sec.~\ref{s:last}.

The fluid equations derived under the assumption of local thermodynamic equilibrium
are called inviscid, or ideal-fluid, equations. 

\subsection{Fluid rest frame}

The rest frame of a fluid element is the galilean frame in which its momentum
vanishes. All thermodynamic quantities associated with a fluid element 
(for example, $\epsilon $, $P$, $n$) 
are defined in the rest frame. They  are therefore Lorentz scalars by construction 
(for the same reason as the mass of a particle is a Lorentz scalar). 
Local thermodynamic equilibrium implies that the fluid element
has isotropic properties in the fluid rest frame. 
This is a very strong assumption. It   will be used extensively below. 
It is, in fact, the only non-trivial assumption of inviscid hydrodynamics. 

\subsection{Fluid velocity}

The velocity $\vec v$ of a fluid element is defined as the 
velocity of the rest frame of this fluid element with respect 
to the laboratory frame. The 4-velocity $u^\mu$ is defined by 
\begin{eqnarray}
u^0&=&\frac{1}{\sqrt{1-\vec v^2}}\cr
\vec u&=&\frac{\vec v}{\sqrt{1-\vec v^2}},
\label{umu}
\end{eqnarray}
where we have chosen a unit system where $c=1$.
$u^0$ is the Lorentz contraction factor. 
The 4-velocity transforms as a 4-vector under Lorentz transformations. 
The square of a 4-vector is a Lorentz scalar, and we indeed obtain
\begin{equation}
u^\mu u_\mu=(u^0)^2-\vec u^2=1.
\label{umuumu}
\end{equation}
In hydrodynamics, the fluid velocity is a function of $(t,x,y,z)$, as
are the thermodynamic quantities $\epsilon$, $P$ and $n$. 
The fluid velocity is also referred to as the ``collective''
velocity.

\subsection{Baryon number conservation}

In nonrelativistic fluid dynamics, the equation of mass conservation is 
\begin{equation}
\frac{\partial\rho }{\partial t}+\vec\nabla(\rho \vec v)=0,
\end{equation}
where $\rho$ is the mass density. 
A relativistic conservation equation must 
take into account the Lorentz contraction of the volume by a factor $u^0$.
Recall that the baryon density $n$ is always defined in the fluid rest frame. 
The baryon density in the moving frame is therefore $n u^0$. Replacing
$\rho $ with $n u^0$ in the above equation and using $\vec u=u^0\vec v$, 
one obtains the following covariant equation:
\begin{equation}
\partial_\mu (n u^\mu)=0,
\label{nconservation}
\end{equation}
where we use the standard notation $\partial_\mu =\partial/\partial x^\mu$. 
This is a conservation equation for the 4-vector $n u^\mu$. 
$n u^0$ is the baryon density, and $n \vec u$ is the baryon flux. 

In the rest frame of the fluid, the baryon flux vanishes. In nonrelativistic
fluid dynamics, this is how the fluid rest frame is defined. In the
relativistic case, the baryon flux could in principle be $\not= 0$ in
the fluid rest frame, defined as the frame where the momentum density is
zero: the momentum of baryons could be compensated by the momentum of
baryonless particles (pions, gluons). 
However, local thermodynamic equilibrium implies isotropy. 
If there was a non-zero current, it would define a direction in space
and isotropy would be lost. The baryon flux therefore vanishes in
inviscid hydrodynamics. 
In relativistic viscous hydrodynamics, which studies deviations from
local thermodynamic equilibrium, the baryon flux may be non-zero in the
local rest frame: this transport phenomenon is called diffusion.

\subsection{Energy and momentum conservation}

The conservation of total energy and momentum gives 4 local conservation equations, 
each of which is analogous to the equation of baryon-number conservation. 
Baryon conservation gives a conserved current, which is a contravariant 4-vector 
$J^\mu=nu^\mu$. Energy and momentum are also a contravariant 4-vector,
therefore the associated conserved currents can be written as a
contravariant tensor $T^{\mu\nu}$,  
where each value of $\nu$ corresponds to a component of the 4-momentum, and each value
of $\mu$ is a component of the associated current.  
Specifically,
\begin{itemize}
\item $T^{00}$ is the energy density
\item $T^{0j}$ is the density of the $j^{th}$ component of momentum, with $j=1,2,3$. 
\item $T^{i0}$ is the energy flux along axis $i$.
\item $T^{ij}$ is the flux along axis $i$ of the $j^{th}$ component of momentum.
\end{itemize}
The momentum flux $T^{ij}$ is usually called the pressure tensor. 
Kinetic pressure is precisely defined as the momentum flux 
(see Sec.~\ref{s:kinetictheory}).

In the fluid rest frame, the assumption of local thermodynamic equilibrium strongly 
constrains the energy-momentum tensor. Isotropy implies that the energy flux $T^{i0}$ 
and the momentum density $T^{0j}$ vanish. In addition, it implies that the pressure 
tensor is proportional to the identity matrix, i.e., $T^{ij}=P\delta_{i,j}$, where 
$P$ is the thermodynamic pressure. The energy-momentum in the fluid rest frame 
is thus
\begin{equation}
T_{(0)}=\left(\begin{array}{cccc}\epsilon &0&0&0\\ 0&P&0&0\\ 0&0&P&0\\ 0&0&0&P\end{array}\right)
\label{tmunu0}
\end{equation}

In order to obtain the energy-momentum tensor in a moving frame, one does a Lorentz 
transformation. In what follows, we shall only need
the expression of $T^{\mu\nu}$ to first order in the fluid velocity. 
To first order in the velocity $\vec v$, the matrix of a Lorentz transformation is
\begin{equation}
\Lambda =\left(\begin{array}{cccc}1 &v_x&v_y&v_z\\ v_x&1&0&0\\ v_y&0&1&0\\ v_z&0&0&1\end{array}\right).
\label{tl}
\end{equation}
Under a Lorentz transformation, the contravariant tensor $T_{(0)}^{\mu\nu}$ transforms to
\begin{equation}
T^{\mu\nu}= \Lambda^{\mu}_{\phantom{m}\alpha}
\Lambda^{\nu}_{\phantom{m}\beta}T_{(0)}^{\alpha\beta},
\end{equation}
which can be written as a multiplication of $(4\times 4)$ matrices
\begin{equation}
T=\Lambda T_{(0)}\Lambda^T,
\label{matrixproduct}
\end{equation}
where $\Lambda^T$ denotes the transpose of $\Lambda$. (\ref{tl}) shows that $\Lambda$ is 
symmetric, $\Lambda^T=\Lambda$. Keeping only terms
to order 1 in the velocity $\vec v$, (\ref{matrixproduct}) gives 
\begin{equation}
T=\left(\begin{array}{cccc}\epsilon &(\epsilon +P)v_x&(\epsilon +P)v_y&(\epsilon +P)v_z\\ 
(\epsilon +P)v_x&P&0&0\\ (\epsilon +P)v_y&0&P&0\\ (\epsilon +P)v_z&0&0&P\end{array}\right)
\label{tmunu1}
\end{equation}

We first note that $T^{\mu\nu}$ is  symmetric: the momentum density $T^{0i}$ 
and the energy flux $T^{i0}$ are equal. This is because Lorentz transformations
preserve the symmetries of tensors, and the tensor of the fluid at rest
(\ref{tmunu0}) is symmetric. The symmetry of $T^{\mu\nu}$ is a nontrivial consequence
of relativity. In nonrelativistic fluid dynamics, the energy
flux and the momentum density differ. (Recall that  
nonrelativistic energy does not include mass energy.) 
They have different dimensions:
the ratio of energy flux to momentum density has the dimension of a velocity squared, 
which is dimensionless in relativity. 

The momentum density is $(\epsilon +P)\vec v$. For a nonrelativistic
fluid, it is $\rho \vec v$, where $\rho $ is the mass density. Since
$P\ll \epsilon $  and $\epsilon \simeq \rho \vec v$ in the
nonrelativistic limit, we recover the correct  
limit. What replaces the mass density for a nonrelativistic fluid is not 
$\epsilon $, as one would naively expect, but $\epsilon +P$: 
pressure contributes to the inertia of a relativistic fluid. 

Finally, we prove that the energy-momentum tensor for an arbitrary fluid velocity is 
\begin{equation}
T^{\mu\nu}=(\epsilon +P)u^\mu u^\nu-P g^{\mu\nu},
\label{tmunu}
\end{equation}
where $g^{\mu\nu}\equiv {\rm diag}(1,-1,-1,-1)$ is the Minkovski metric tensor.
One easily checks that this equation reduces to (\ref{tmunu0}) in the rest frame of the 
fluid, where  $u^\mu=(1,0,0,0)$. In addition, 
both sides of (\ref{tmunu}) are contravariant tensors, which means that they transform
identically under Lorentz transformations. Since they are identical in one frame, they 
are identical in all frames, which proves the validity of (\ref{tmunu}).

The conservation equations of energy and momentum are 
\begin{equation}
\partial_\mu T^{\mu\nu}=0.
\label{dmutmunu}
\end{equation}
(\ref{nconservation}), (\ref{tmunu}) and (\ref{dmutmunu}) are the equations of 
inviscid relativistic hydrodynamics. Together with the equation of state of the fluid, 
which is defined as a functional relation between $\epsilon $, $P$ and $n$, they form 
a closed system of equations. 

For sake of simplicity, only continuous flows will be studied, in
which all quantities vary continuously with space-time coordinates. 
Inviscid
hydrodynamics has a whole class of discontinuous solutions, which are
called ``shock waves''. 
The entropy of the fluid increases through a shock, while 
it is constant for a continuous flow (see Sec.~\ref{s:cooling}, and
problem 1 in appendix). 
Shock waves usually appear when the fluid undergoes compression, not
expansion. They are therefore of limited relevance to heavy-ion
collisions~\footnote{In fact, shock waves do appear in the expansion
  when the equation of state has a first-order phase
  transition. These ``rarefaction shocks'' produce little entropy, at
  most 7\%~\cite{friman}.}.  

\subsection{Sound waves}

Sound is defined as a small disturbance propagating in a uniform fluid at rest.
The energy density and pressure are written in the form
\begin{eqnarray}
\epsilon (t,x,y,z)&=&\epsilon _0+\delta \epsilon(t,x,y,z) \cr
P(t,x,y,z)&=&P_0+\delta P(t,x,y,z),
\label{soundexp}
\end{eqnarray}
where $\epsilon _0$ and $P_0$ correspond to the uniform fluid, 
and $\delta \epsilon $ and $\delta P$ correspond to the small disturbance.
To study the evolution of this disturbance, we linearize the equations
of 
energy-momentum conservation by keeping only terms up to first order in 
$\delta \epsilon $, $\delta P$ and $\vec v$. For this purpose, the expression
(\ref{tmunu1}) will suffice, since it is correct to first order in the velocity.
(\ref{dmutmunu}) gives 
\begin{eqnarray}
\frac{\partial\epsilon }{\partial t}+\vec\nabla\cdot((\epsilon +P)\vec v)&=&0\cr
\frac{\partial}{\partial t}((\epsilon +P)\vec v)+\vec\nabla P&=&\vec 0.
\label{firstordereqs}
\end{eqnarray}
Inserting (\ref{soundexp}) and linearizing, these equations simplify:
\begin{eqnarray}
\frac{\partial(\delta \epsilon) }{\partial t}+(\epsilon_0 +P_0)\vec\nabla\cdot\vec v&=&0\cr
(\epsilon_0 +P_0)\frac{\partial\vec v}{\partial t}+\vec\nabla \delta P&=&\vec 0.
\label{sound1}
\end{eqnarray}
The first equation expresses that the density decreases if the velocity field diverges, 
$\vec\nabla\cdot\vec v>0$, i.e., if the volume increases. This is energy conservation.
The second equation is Newton's second law, that the inertia of the fluid multiplied by 
its acceleration must be equal to the force. The force per unit volume is 
$-\vec\nabla P$. It pushes the fluid towards lower
pressure. 

We now define the velocity of sound $c_s$ by:
\begin{equation}
c_s=\left(\frac{\partial P}{\partial \epsilon }\right)^{1/2}.
\label{defcs}
\end{equation}
$c_s^2$ is inversely proportional to the compressibility of the
fluid. A ``soft'' equation of state corresponds to a small $c_s$. 
The derivative in (\ref{defcs}) is well defined only if we specify
along which line the partial derivative is taken. 
It will  be shown in problem 1 that in ideal fluid dynamics, the
entropy per baryon  
of a fluid element is conserved as a function of time. If the fluid is initially uniform,
then the entropy per baryon remains constant throughout the fluid at
all times.  
This means that the partial derivative must be taken along the lines of constant
entropy per baryon, $s/n$ (thus corresponding to the adiabatic
compressibility). In the case of a baryonless quark-gluon  
plasma, there is only one degree of freedom, and no ambiguity in
defining the derivative. Using (\ref{dP}) and (\ref{depsilon}), 
one can rewrite $c_s$ as
\begin{equation}
c_s=\left(\frac{d\ln T}{d\ln s}\right)^{1/2}
\label{csbis}
\end{equation}
for a baryonless fluid. 

Using the definition (\ref{defcs}), we write $\delta P=c_s^2 \delta \epsilon $ in 
(\ref{sound1}). We then eliminate $\vec v$ between the two equations:
\begin{equation}
\frac{\partial^2(\delta \epsilon) }{\partial t^2}-c_s^2 \Delta (\delta \epsilon)=0.
\label{sound2}
\end{equation}
This is a wave equation in 3+1 dimensions, with velocity $c_s$. This equation 
means that small perturbations in a uniform fluid travel at the
velocity $c_s$, independent of the shape of the perturbation: there is
no sound dispersion in an inviscid fluid. 

\subsection{Ideal gas}

If the interaction energies between the particles are small compared
to their kinetic energies, one can express the hydrodynamic quantities
in terms of the individual particle properties: conserved baryon
number $B$, 
velocity $\vec v$ and momentum  $p^\mu$. We use the notation 
$v^\mu$ for $(1,\vec v)$, or equivalently, $v^\mu=p^\mu/p^0$. 
Please note that in spite of the notation,
$v^\mu$ does not transform like a 4-vector under a Lorentz boost. 
The baryon current and energy-momentum tensor of a small fluid element
of volume $V$ are 
\begin{eqnarray}
n u^\mu&=&\frac{1}{V}\sum_{\rm particles} B v^\mu\cr
T^{\mu\nu}&=&\frac{1}{V}\sum_{\rm particles} p^\nu v^\mu .
\label{tmunuideal}
\end{eqnarray}
With these definitions, it is straightforward to check that $n u^0$,
$T^{00}$ and $T^{0i}$ correspond to the baryon, energy and
momentum densities, respectively. The corresponding fluxes are obtained 
by weighting these quantities with the particle velocity $\vec v$. 

Using the assumption of local thermodynamic equilibrium, one can
replace these quantities with their thermal averages. The average number
of particles with momentum $\vec p$ is given by Boltzmann statistics 
(we neglect quantum statistics for simplicity), (\ref{maxwell}),
where we replace $E_{\vec p}$ with the energy in the fluid rest frame $E^*$. 
Using (\ref{riemannsum}), one can do the following substitution:
\begin{equation}
\frac{1}{V}\sum_{\rm particles}
\to\int \frac{d^3 p}{(2\pi \hbar)^3}
e^{(-E^*+\mu)/T}.
\end{equation}
This result will be useful later. 
We finally show that the expressions in (\ref{tmunuideal}) are
covariant. For this purpose, we write $v^\mu=p^\mu/p^0$, and we note
that $d^3 p/p^0$ is a Lorentz scalar, so that $n u^\mu$ and $T^{\mu
  \nu}$ are explicitly covariant. 

\section{Hydrodynamical expansion}
\label{s:heavyions}

The energy of a nucleus-nucleus collision at RHIC is 100 GeV per nucleon. 
This means that each incoming nucleus is contracted by a Lorentz factor 
$\gamma \approx 100$: nuclei are thin pancakes colliding. The
collision creates thousands of particles in a small volume. These
particles interact.  
If these interactions are strong enough, the system may reach a state
of local thermodynamic equilibrium. Equilibrium is at best local, certainly
not global: global equilibrium applies to a gas in a closed box, which stays
there for a long time and becomes homogeneous. The system formed in a
heavy-ion collision starts expanding as soon as it is produced, and is
far from homogeneous.  

Can QCD tell whether or not the system reaches thermodynamic
equilibrium? There is not yet an answer to this question, but a lot of
progress has been made on this issue in recent years, due in
particular to works on QCD plasma instabilities~\cite{Arnold:2003rq}. 
Another question is: can we tell from experimental data 
whether the system has reached local equilibrium? This issue will be 
briefly touched upon in Sec.~\ref{s:last}. You should keep in mind that 
local equilibrium is, at best, an approximation. Even if it turns 
out to give reasonable results, it is not the end of the story.

In this section, we assume that the system of interacting fields and 
particles produced in the collision reaches local 
thermodynamic equilibrium at some point. Its subsequent evolution 
follows the laws of inviscid hydrodynamics. Since there are first-order 
partial differential equations, their solution is uniquely determined 
once initial conditions are specified, together with an equation of state.

\subsection{Initial conditions}
\label{s:initial}

The $z$-axis is chosen as the collision axis, and the origin is chosen
such that the collision starts at $z=t=0$. The two nuclei pass through 
each other in a time $t_{\rm coll}\sim 0.15$~fm/c at RHIC. This time is 
a factor 100 smaller than the other characteristic dimension, the 
transverse size $R$ of the nucleus. This clear hierarchy between the 
two scales is crucial.

The initial conditions are fixed at some initial time $t_0$
(or more generally, on a space-like hypersurface). A complete set 
of initial conditions involves the 3 components of fluid velocity,
the energy density and the baryon density, at each point in space.

If the thermalization time $t_0$ is short enough, the transverse
components $v_x$ and $v_y$ of the fluid velocity are zero. The reason
is that the parton-parton collisions which produce particles occur on 
very short transverse scales. They produce particles whose transverse
momenta are distributed isotropically in the transverse plane. 
Isotropy implies that there is no preferred direction, and that
the transverse momentum averaged over a fluid element vanishes. 
This part of the initial conditions is the only one on which there
is fairly general agreement.  
This is the reason why the clearest experimental signatures of
hydrodynamic behaviour are those associated with ``transverse flow'', 
as we shall see below:
if there is no transverse collective motion initially present in 
the system and if we see it in the data, it means that something 
has happened inbetween which has to do with hydrodynamics.

We now discuss the initial value of the longitudinal flow velocity
$v_z$.  All particles are produced in a very short interval
around $z=t=0$. The standard prescription is that their longitudinal
motion is uniform, so that their velocity is $v_z=z/t$: all particles
at a given $z$ have the same $v_z$, hence it is also the fluid
velocity. This prescription is boost invariant,  
in the following sense: if one does a homogeneous\footnote{A Lorentz
  transformation is homogeneous if it leaves the origin unchanged.} 
Lorentz transformation 
with a velocity $v$ along the $z$ axis, all three quantities
$v_z$, $z$, $t$ are transformed, but $v_z=z/t$ still holds in the 
new frame. This is because uniform motion remains uniform under a 
Lorentz transformation. This ``boost-invariant'' prescription
was first proposed by Bjorken~\cite{Bjorken:1982qr}, and it is
supported by models inspired by high-energy QCD, such as the
colour glass condensate. 

\begin{figure}
\begin{center}
\includegraphics*[width=0.55\linewidth]{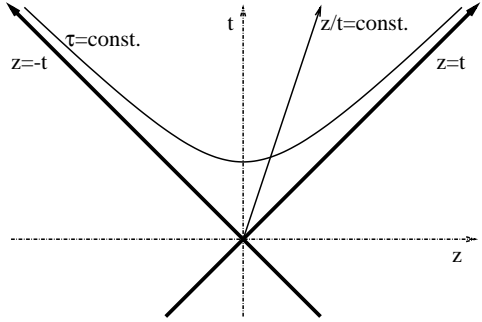}
\end{center}
\caption{Nucleus-nucleus collision in the $(z,t)$ plane. The
thick lines are the trajectories of the colliding nuclei, which
are moving nearly at the velocity of light. The lines of constant
$z/t$ are also lines of constant space-time rapidity $ \eta_s $. 
\label{fig:bjorken}}
\end{figure}

We now introduce new coordinates, 
the proper time $\tau$, the space-time rapidity $ \eta_s $,
and the fluid rapidity $Y$, defined by: 
\begin{eqnarray}
t&=&\tau\cosh\eta_s\cr
z&=&\tau\sinh\eta_s\cr
v_z&=&\tanh Y. 
\label{deftaueta}
\end{eqnarray}
Under a Lorentz boost in the $z$ direction, $\tau$ is unchanged, while 
$\eta_s$ and $Y$ are shifted by a constant. 
Lines of constant $\tau$ and constant $ \eta_s $ are represented in
figure~\ref{fig:bjorken}. 
Initial conditions are usually specified at a given proper time 
$\tau=\tau_0$, rather than at a given time $t=t_0$. 
Bjorken's prescription $v_z=z/t$ translates into $Y=\eta_s$: the fluid
rapidity equals the space-time rapidity.

We now discuss the initial density profile. One usually specifies the
energy density, or the entropy density, as a function of
$x,y,\eta_s$. 
There are 
constraints on these profiles, both theoretical and experimental, and
prescriptions which satisfy these constraints. 
On the theoretical side, there is locality: it implies that a given 
point $(x,y)$ in the transverse plane, the initial density can depend
only on the thickness functions $T_A$ and $T_B$  of the two colliding  
nuclei at this point, defined as the integrals 
\begin{equation} 
T_{A,B}(x,y)=\int_{-\infty}^{+\infty} \rho_{A,B}(x,y,z)dz,
\end{equation} 
where $\rho_A(x,y,z)$ (resp. $\rho_B$) is the density of nucleons per
unit volume in nucleus $A$ (resp. B). The initial energy density is 
$\epsilon(x,y,\eta_s)=f(T_A(x,y),T_B(x,y),\eta_s)$, where $f$ is some function. 
Various prescriptions can be found in the literature
\begin{itemize}
\item The initial energy density is proportional to the density of
  binary collisions $T_A T_B$~\cite{Kolb:2003dz}. 
\item The initial entropy density is proportional to the density of
  participants~\cite{Ollitrault:1991xx}, 
which is essentially  $T_A+T_B$ on the overlap area,   and 0 outside. 
More complex prescriptions can also be found, where the 
  entropy~\cite{Hirano:2004ep} or the energy~\cite{Nonaka:2006yn} density
 are linear combinations of the densities of binary collisions and
  participants. 
\item In contrast to the above pictures, where the $\eta_s$ dependence
  is fitted to measured rapidity spectra, the colour glass
  condensate~\cite{Hirano:2004en} provides a prescription for the 
  $\eta_s$ dependence. It also predicts quite distinctive transverse
  profiles: at $z=\eta_s=0$, it gives an initial multiplicity density
  approximately proportional to $\min (T_A,T_B)$ \cite{Lappi}.
\end{itemize}
All these prescriptions reproduce well the observed centrality
dependence of the global multiplicity.

\begin{figure}
\begin{center}
\includegraphics*[width=0.45\linewidth]{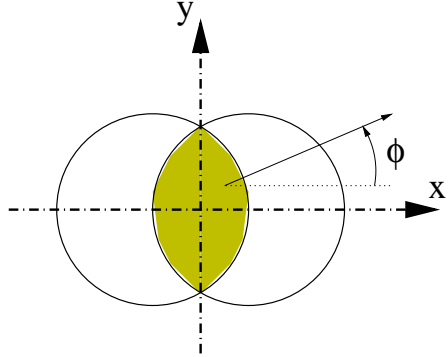}
\end{center}
\caption{Non-central nucleus-nucleus collision in the transverse
  $(x,y)$ plane. The $x$-axis is chosen as the direction of the impact
  parameter. The shaded area is the overlap area between the nuclei, 
where particles are produced. The density in this area can be
 approximated by a gaussian, (\ref{gaussian}). The azimuthal
  angle of an outgoing particles with the $x$-axis is denoted by
  $\phi$. 
\label{fig:xy}}
\end{figure}

We assume 
for simplicity a gaussian entropy density profile at $\tau=\tau_0$:
\begin{equation}
s(x,y,\eta_s)\propto \exp\left(-\frac{x^2}{2\sigma_x^2}
-\frac{y^2}{2\sigma_y^2}-\frac{\eta_s^2}{2\sigma_\eta^2}\right).
\label{gaussian}
\end{equation}
In this equation, $\sigma_x$ and $\sigma_y$ are the rms (root mean
square) widths of the transverse distribution. For a central Au-Au
collision, $\sigma_x=\sigma_y\simeq 3$~fm. 
For a non-central collision, one chooses in general the
$x$-axis as the direction of impact parameter (see figure~\ref{fig:xy}),
and $\sigma_x<\sigma_y$. For a Au-Au collision at impact parameter 
$b=7$~fm, $\sigma_x\simeq 2$~fm, $\sigma_y\simeq 2.6$~fm. 
Unlike $\sigma_x$ and $\sigma_y$, $\sigma_\eta$ is 
dimensionless. 
In order to estimate its  value, we use the fact that
the particle  multiplicity is proportional to the entropy. We further
assume, for sake of simplicity, that  the rapidity of outgoing
particles is equal to their space-time rapidity $\eta_s$. 
Rapidity distributions of 
outgoing particles in symmetric nucleus-nucleus collisions at RHIC are
perfectly  fit by gaussians of width $\sigma_\eta\simeq
2.3$~\cite{Bearden:2004yx}.

\subsection{Longitudinal acceleration}

Our initial condition for the longitudinal
fluid velocity is $v_z=z/t$ or, equivalentely, $Y=\eta_s$. 
The original Bjorken picture~\cite{Bjorken:1982qr} is that 
this relation holds at all times. 
We now discuss under which this condition it is 
preserved by the hydrodynamical evolution. 
We first study the simple case $z=0$. Since $v_z=0$ initially at
$z=0$, we can use (\ref{firstordereqs}), which is valid to first order
in the fluid velocity. 
The $z$ component gives:
\begin{equation}
\frac{\partial}{\partial t}((\epsilon +P)v_z)+\frac{\partial}{\partial
  z}P=0.
\end{equation}
Recalling that $v_z=0$ initially, we obtain the acceleration:
\begin{equation}
\frac{\partial v_z}{\partial
  t}=-\frac{1}{\epsilon+P}\frac{\partial P}{\partial z}
=-c_s^2\frac{\partial \ln s}{\partial z}, 
\label{zaccel}
\end{equation}
where we have used (\ref{isentropic}) and (\ref{defcs}). 
If the initial density $s$ depends on $z$, the fluid is accelerated,
and the initial condition $v_z=0$ is not preserved by the 
hydrodynamical evolution.

We now rewrite 
(\ref{zaccel}) using the variables $\eta_s,\tau,Y$ defined in 
(\ref{deftaueta}). 
Near $z=0$, $dt\simeq d\tau$, $dz\simeq \tau d\eta_s$, 
$v_z\simeq Y$:
\begin{equation}
\frac{\partial Y}{\partial\tau }
=-\frac{c_s^2}{\tau}\frac{\partial \ln s}{\partial \eta_s}.
\label{zaccel2}
\end{equation}
This equation can easily be generalized to $\eta_s\not= 0$ using boost
invariance: any 
value of $z$ with $|z|<t$ can be brought to $z=0$ by means of a
homogeneous Lorentz boost in the $z$ direction. 
Such a boost leaves $\tau$ unchanged, and shifts $ \eta_s $ by a 
constant. Hence it leaves $(\partial/\partial \eta_s )_\tau$ unchanged, 
and (\ref{zaccel2}) holds for all $\eta_s$.
The general condition under which $Y=\eta_s$ at all times is
$(\partial s/\partial  \eta_s )_\tau=0$, 
corresponding to the limit $\sigma_\eta\to\infty$ in (\ref{gaussian}),
i.e., to flat rapidity spectra.  

We now show that even though rapidity spectra are not flat, the Bjorken 
picture is a reasonable approximation in practice at high energies. We
follow the same approach as Eskola {\it et
  al\/}~\cite{Eskola:1997hz}. 
(\ref{gaussian}) and (\ref{zaccel2}) give 
\begin{equation}
\frac{\partial Y}{\partial\tau}=\frac{c_s^2}{\tau}\frac{\eta_s}{\sigma_\eta^2}.
\end{equation}
Integrating from $\tau_0$ to $\tau$ with the initial condition
$Y=\eta_s$, we obtain our final result
\begin{equation}
Y(\tau)=\left(1+\frac{c_s^2 \ln(\tau/\tau_0)}{\sigma_\eta^2}\right) \eta_s.
\label{zaccel4}
\end{equation}
The rapidity of the fluid is not {\it equal\/} to the initial
rapidity, as assumed in the Bjorken scenario, but {\it proportional\/}
to it. As will be explained in Sec.~\ref{s:timescale}, 
transverse expansion acts as a cutoff for the longitudinal expansion
at a time $\tau\sim 3.6$~fm/c for a central Au-Au collision. 
Even if the longitudinal pressure builds up as early as
$\tau_0\simeq 1$~fm/c, and assuming $c_s^2=\frac{1}{3}$, this results in a
modest increase of the rapidity width, by 9\%. At LHC energies, where
the rapidity width $\sigma_\eta$ is expected to increase, effects of
longitudinal acceleration will be even smaller.

\subsection{Longitudinal cooling}
\label{s:cooling}

We now derive the evolution of baryon density, energy density and
entropy density in the Bjorken picture of a uniform longitudinal
expansion. We assume that the transverse components of the 
velocity, together with their spatial derivatives, remain
negligible. As will be 
shown in Sec.~\ref{s:timescale}, this is a good approximation as long as
$t\ll \sigma_x/c_s,\sigma_y/c_s$. 

We start with the baryon density. We rewrite (\ref{nconservation})
at $z=0$. The Bjorken
prescription $v_z=z/t$ gives $v_z=0$ and $\partial v_z/\partial
z=1/t$:
\begin{equation}
\frac{\partial n}{\partial t}+\frac{n}{t}=0.
\label{longexpn}
\end{equation}
This equation can be integrated as $nt$=constant, which expresses the
conservation of baryon number in a comoving fluid element: 
neglecting the transverse expansion, the volume of a comoving fluid
element increases like the longitudinal size, i.e., like $t$. 

The evolution of energy density at $z=0$ is derived in a similar way, using
(\ref{firstordereqs}):
\begin{equation}
\frac{\partial\epsilon }{\partial t}+\frac{\epsilon +P}{t}=0.
\label{longexp}
\end{equation}
The generalization of (\ref{longexpn}) and (\ref{longexp}) for
arbitrary $z$ is obtained by transforming to $(\tau, \eta_s )$
coordinates, and replacing $(\partial/\partial t)_z$ with
$(\partial/\partial\tau)_{\eta_s} $, and $t$ with $\tau$. 

Unlike the baryon number, the total energy in a comoving fluid element
is not conserved. (\ref{longexp}) can be recast in the form 
\begin{equation}
d(\epsilon t)=-Pdt,
\label{pdv}
\end{equation}
which shows that the comoving energy decreases. This is due to 
the negative work of pressure forces, $dE=-PdV$. 
This result is by no means trivial. It relies on our assumption of
local equilibrium, which implies that the pressure is isotropic. $P$
in (\ref{pdv}) comes from $T^{33}$ in (\ref{tmunu0}), i.e., it
is really the longitudinal pressure. For an ideal gas,
(\ref{tmunuideal}) shows that $T^{33}=\sum_{\rm particles}p_z v_z$. If the
particles are initially produced with $v_z=z/t$ (as is for instance
the case in the colour glass condensate), the longitudinal pressure
vanishes at $z=0$. 
A non-zero longitudinal pressure can only appear as
a result of the thermalization process. Most of the work on
thermalization is about understanding how this longitudinal pressure
appears. 

It is worth noting that there is no direct evidence for longitudinal
cooling, (\ref{pdv}), from experimental data. Experimental data are
particles, which are emitted mostly at the final stage of the
evolution. Our knowledge of initial stages is indirect. 
Longitudinal cooling implies a higher initial energy, for a given
final energy. This can be observed only through a direct signature of 
the initial temperature. The most promising observables in this 
respect are electromagnetic observables, ``thermal'' dileptons and 
photons, which are mostly emitted at the early stages, and sensitive
to the temperature, but they are plagued by huge backgrounds. 
Although there is no experimental evidence for longitudinal cooling,
it is clearly favoured theoretically:
models of particle production based on perturbative
QCD produce an initial energy significantly higher than the final
energy, typically by a factor of 3~\cite{Eskola:2005ue,Eskola:2005uf},
and require substantial longitudinal cooling to match with the data. 

Finally, it is worth noting that the total entropy of a comoving fluid
element is conserved, as the baryon number: (\ref{dE}) shows that
$dE=-PdV$ and $dN=0$ implies $dS=0$.  This is a general result for 
inviscid hydrodynamics 
(see problem 1 in appendix). Physically, it means that there is no
heat diffusion between fluid cells. 
To show this explicitly, we multiply (\ref{longexpn}) by $\mu$ and
subtract (\ref{longexp}). Using (\ref{Extensive3}) and
(\ref{depsilon}), one obtains  
\begin{equation}
\frac{\partial s}{\partial t}+\frac{s}{t}=0, 
\end{equation}
which shows that the comoving entropy, which scales like $st$, is constant.

\subsection{Orders of magnitude}
\label{s:orders}

We can use experimental data to estimate the initial density in a
heavy-ion collision. A popular estimate is Bjorken's estimate 
of the energy density~\cite{Bjorken:1982qr}, defined as the ratio of
the final ``transverse'' energy (defined as $E\sin\theta$, 
where $\theta$ is the relative angle between the particle velocity and the
collision axis, or polar angle) to the initial volume. This estimate
neglects the longitudinal cooling (\ref{pdv}), and therefore
underestimates the initial energy density. 

It is in fact probably safer to assume that the number of particles
remains constant throughout the evolution: in the quark-gluon plasma
phase, the particle number is approximately 
proportional to entropy (see (\ref{entropyperparticle})), and entropy
is conserved. In the hadron phase, the scenario of chemical freeze-out 
(see below Sec.~\ref{s:freezeouts}) supports particle number
conservation in the hadronic phase. Finally, at the quark-hadron phase 
transition, the idea of local quark-hadron duality (taken over from
perturbative QCD~\cite{Azimov:1984np,Melnitchouk:2005zr}) suggests
that the number 
of particles might again be conserved. It is interesting to note that while
perturbative QCD estimates fail in calculating the energy, they give a
gluon multiplicity comparable to the observed
multiplicity~\cite{Eskola:2005ue}, which seems to support this 
assumption. 

In order to estimate the initial density, we assume for simplicity
that the longitudinal velocity of particles remains constant, i.e., 
$v_z=z/t$. Then, the particle density at time $t$ is 
\begin{equation}
n(t)=\frac{1}{S}\frac{dN}{dz}=\frac{1}{St}\frac{dN}{dv_z},
\end{equation}
where $S$ it the transverse area of the interaction region, 
$S\approx \pi R^2\approx 100$~fm$^2$ for a central Au-Au collision,
and $N$ is the particle multiplicity. Since we are interested in the
particle density in the 
fluid rest frame, we choose to estimate it near $z=0$, where the fluid
is at rest.  
The PHOBOS collaboration has measured~\cite{Back:2002uc}  the polar
angle distribution of charged particles in central Au-Au collisions at
100 GeV per nucleon~\footnote{What is measured is in fact the 
  pseudorapidity ($\eta$)  distribution, defined by
  $dN/d\eta=\sin\theta\, 
  dN/d\theta$, which coincides with the polar-angle distribution near
  $\theta=\pi/2$.}. The result is $dN_{\rm ch}/d\theta\simeq 700$ at
$\theta=\pi/2$. 
Now, $v_z=v\cos\theta$. For particles emitted near $\theta=\pi/2$ with
velocity $v$, this gives $dN/dv_z=(1/v)dN/d\theta$. 
The factor $(1/v)$ gives on average a factor 1.25, and charged
particles are only 2/3 of the produced particles, so that
$dN/dv_z\simeq 1300$. 
This gives numerically, for a central Au-Au collision at the top RHIC
energy, 
\begin{equation}
n(t)\simeq\frac{13}{t}, 
\label{RHICdensity}
\end{equation}
where $n$ is in fm$^{-3}$ and $t$ in fm/c. 

This estimate must be compared with our estimate of the particle
density in a quark-gluon plasma, (\ref{idealqgp}). Lattice QCD
predicts that the transition to the quark-gluon plasma occurs near 
$T_c\approx 192$~MeV~\cite{Karsch:2007vw}. Since $\hbar
c=197$~MeV$\cdot$fm, and we have 
chosen $c=1$ throughout the calculations, (\ref{idealqgp}) gives
\begin{equation}
n\simeq 3.75\, {\rm fm}^{-3}.
\end{equation}
at $T_c$. Comparing with (\ref{RHICdensity}), one sees that the system
is above the critical density only if $t<3.5$~fm/c: the lifetime of
the quark-gluon plasma is approximately 3.5~fm/c.  
This is of course a rough estimate: the density profile is not
homogeneous throughout the surface $S$ (the maximum density, at the
center, is approximately twice larger than the average density, and
the lifetime is correspondingly larger), and we have neglected the
transverse expansion (which, on the contrary, reduces the lifetime). 
Finally, our starting assumption that the particle number is conserved 
is a crude picture for two reasons: The number of particles
is ill-defined in a strongly-interacting system. Recent works have
argued that the hadronization process could involve both fragmentation
and recombination of partons, thus breaking the conservation of
particle number at the transition~\cite{Fries:2003kq}.

\subsection{The onset of transverse expansion}
\label{s:onset}

The initial transverse velocity of the fluid is 0, but the
acceleration is in general not zero. It is given by an equation 
similar to (\ref{zaccel}):
\begin{equation}
\frac{\partial v_x}{\partial  t}=-\frac{1}{\epsilon+P}\frac{\partial
  P}{\partial x} =-c_s^2\frac{\partial
 \ln s}{\partial x}.
\label{xaccel}
\end{equation}
and a similar equation along the $y$-axis. 
Inserting (\ref{gaussian}) into
(\ref{xaccel}), and assuming constant $c_s$ for simplicity, we
integrate over $t$ to obtain, for small $t$,
\begin{equation}
v_x=\frac{c_s^2 x}{\sigma_x^2}t,\ \ \ \ 
v_y=\frac{c_s^2 y}{\sigma_y^2}t.
\label{vxy}
\end{equation}
Note that we have integrated from $t=0$. Thermalization
certainly requires some time, and hydrodynamics cannot apply at very
early times. On the other hand, the system is expanding freely in the
vacuum, and it is clear that the transverse expansion starts 
immediately: it does not wait until thermalization is achieved, so 
that it is probably reasonable to start the transverse expansion at
$t=0$. 

(\ref{vxy}) shows that
the transverse expansion, unlike the longitudinal expansion, is a very 
smooth process. 
This may not be intuitive: the pressure is very
high at early times, and pressure gradients are largest too, so that a 
huge force $-\vec\nabla P$ acts on the system; but this is
compensated by the large inertia  
$\epsilon+P$, resulting in a linear increase of the transverse fluid 
velocity. 
The typical timescale for transverse expansion is
$\sigma_x/c_s$, which means that longitudinal expansion dominates for 
$t\ll \sigma_x/c_s$. 

The almond shape of the overlap area, in a non-central collision 
(see figure~\ref{fig:xy}) results in $\sigma_x<\sigma_y$, which in
turn implies $\langle v_x^2\rangle>\langle v_y^2\rangle$, where
angular brackets 
denote averages weighted with the initial density:
the tranverse expansion is larger along the smaller dimension, because
the pressure gradient is 
larger. This results in more particles emitted near $\phi=0$ and
$\phi=\pi$, i.e., parallel to the $x$-axis, than near $\phi=\pm
\pi/2$, parallel to the $y$-axis~\cite{Ollitrault:1992bk}. 
This effect corresponds to a $\cos 2\phi$ term in the Fourier
decomposition of the azimuthal distribution:
\begin{equation}
\frac{dN}{d\phi}\propto 1+2 v_2\cos 2\phi.
\label{defv2}
\end{equation}
where $v_2$ is a positive coefficient, which is called ``elliptic
flow''. The observed dependence of $v_2$ on transverse momentum and
particle species is considered the most solid evidence for
hydrodynamical behaviour in nucleus-nucleus collision. It will be studied
in Sec.~\ref{s:elliptic}.  

\subsection{The time scale of transverse expansion} 
\label{s:timescale}

Our equation for longitudinal cooling, (\ref{pdv}), was derived
neglecting transverse expansion. If there was no transverse expansion,
the system would cool forever and no energy would be left in the
central rapidity region. Transverse expansion effectively acts as a
cutoff for longitudinal cooling. The typical time when transverse
expansion becomes significant is, for dimensional reasons, 
$\sigma_x/c_s$ or $\sigma_y/c_s$. 
A convenient scaling variable is provided by the following 
quantity~\cite{Bhalerao:2005mm}:
\begin{equation}
R\equiv \left(\frac{1}{\sigma_x^2}+\frac{1}{\sigma_y^2}\right)^{-1/2}.
\end{equation}
The total transverse energy can be computed, to a very good
approximation, by assuming that (\ref{pdv}) holds until $t=R/c_s$,
and that the energy remains constant for
$t>R/c_s$~\cite{Ollitrault:1991xx}. 
This is what I mean by saying that transverse expansion acts 
as a cutoff for longitudinal cooling. 

An important feature of hydrodynamical models is that the momentum
distributions of outgoing particles depend on the equation of state,
therefore experimental data constrain the equation of state.
Most of this dependence is a consequence of the simple picture above:
after $t=R/c_s$, the energy and entropy of the
fluid are essentially constant. Since the multiplicity is proportional
to the entropy, this also implies that the average energy per particle
remains constant. The transverse energy per particle 
thus reflects the thermodynamic state of the system at $t\approx R/c_s$. 
Since the energy per particle scales like the temperature (see
(\ref{idealqgp})), it gives a direct information on the
temperature of the system at $t\approx R/c_s$. The entropy density at
this time is proportional to the particle density, derived in 
Sec.~\ref{s:orders}.  
Experimental data imply a low temperature, which in turn means that
the equation of state is ``soft'' (see
(\ref{csbis})). Hydrodynamical models favour a soft
equation of state, even softer than predicted by lattice QCD.

Quite naturally, $R/c_s$ is also the characteristic time for
the build-up of elliptic flow: $v_2$ at $t=R/c_s$ is typically half 
its final value. With $c_s=1/\sqrt{3}$, the numerical value of $R/c_s$
for a Au-Au collision  is 3.6~fm/c for $b=0$ (central collision),
2.7~fm/c for $b=7$~fm. 
This explains why elliptic flow is considered a signature of early
pressure. 

The final value of elliptic flow is a good illustration of how the
choice of initial conditions may influence the results. 
Early hydrodynamical calculations had predicted a $v_2$ as large as
seen as RHIC, and this was the main reason for the success of inviscid
hydrodynamics. However, the possibility was raised recently that this 
agreement might be due to unrealistic initial conditions. Let us
briefly explain why. Hydrodynamics predicts that $v_2$ is 
proportional to the eccentricity $\varepsilon$ of the initial
distribution, defined as 
\begin{equation}
\varepsilon\equiv\frac{\sigma_y^2-\sigma_x^2}{\sigma_y^2+\sigma_x^2}. 
\end{equation}
Early hydrodynamical calculations estimated $\varepsilon$ using
participant scaling, or binary collision scaling (see
Sec.~\ref{s:initial}). 
It was discovered recently that the colour glass condensate predicts a
significantly higher
eccentricity~\cite{Hirano:2004ep,Drescher:2006pi}. 

Another effect may increase the initial eccentricity, and was
suggested by experimental data: one expects $\varepsilon$ to
vanish for central collisions, but experimentally, a non-zero $v_2$
is seen even for the most central collisions. Surprisingly, 
the effect is 
larger with smaller nuclei: the value of $v_2$ in central Cu-Cu
collisions is almost twice as large as in central Au-Au collisions. 
The PHOBOS collaboration has suggested that this may be due to 
fluctuations in the positions of nucleons within the
nuclei~\cite{Socolowski:2004hw,Manly:2005zy,Manly2}. There have been several
attempts by STAR 
and PHOBOS to measure these fluctuations directly, but they are
difficult to isolate from other effects. The present situation is that
our knowledge of the initial density profile is much more uncertain
than was usually thought a few years ago.

\section{Particle spectra and anisotropies}
\label{s:spectra}

The fluid eventually becomes free particles which reach the
detector. In this section, we derive some properties of the 
momentum distribution of particles emitted by a fluid.
The transition between a fluid (where the particles undergo many
collisions) and free particles cannot be described by fluid mechanics 
itself. If inviscid hydrodynamics holds throughout most of the 
expansion, one can reasonably assume that the late stages of the
expansion do not alter the essential features of the 
momentum distributions. We therefore assume that the momentum
distribution of outgoing particles is essentially the momentum 
distribution of particles within the fluid, towards the end of the
hydrodynamical expansion, and that the fluid consists of independent
particles (ideal gas). These assumptions form the basis of the
common ``Cooper-Frye freeze-out picture''~\cite{Cooper:1974mv}. 
Here, we further assume that the fluid is 
baryonless, and that momentum distributions are given by 
Boltzmann statistics: 
\begin{equation}
\frac{dN}{d^3 x d^3
  p}=\frac{2S+1}{(2\pi\hbar)^3}\exp\left(-\frac{E^*}{T}\right), 
\label{cooperfrye}
\end{equation}
where $2S+1$ is the number of spin degrees of freedom, and $E^*$ is
the energy of the particle in the fluid rest frame.  
$T$ is called the freeze-out temperature. 

\subsection{Comoving particles and fast particles}

The Boltzmann factor (\ref{cooperfrye}) is maximum when the energy 
$E^*$ in the fluid rest frame is minimum. 
For a given fluid velocity, $E^*$ is minimum when the particle is at
rest in the fluid rest frame, in which case $E^*=m$. This in turn
means that the particle 
velocity in the laboratory equals the fluid velocity: the particle is
comoving with the fluid, and has a momentum $p^\mu=m u^\mu$. For light
particles, this corresponds to 
low transverse momenta: even if the fluid has a transverse velocity as
large as 0.7, the corresponding transverse momentum is approximately 
equal to the mass, i.e., only 140~MeV/c for pions, 500~MeV/c for
kaons. 
In this low-momentum region, the momentum distribution depends 
on how the fluid velocity is distributed, and few
general results can be obtained. 

In this section, we study particles which move faster than the fluid,
which we call ``fast particles''.
For fast particles, $E^*$ is larger than $m$. For a given momentum 
$\vec p$ of the particle, the minimum of $E^*$ occurs if the 
fluid velocity is parallel to $\vec p$: fast particles 
are more likely to be emitted from regions where the fluid velocity is
parallel to their velocity (which means that the fluid and the
particle have the same azimuthal angle $\phi$ and rapidity $y$). This
result can be justified rigorously 
using the saddle-point method~\cite{Borghini:2005kd}. 
For simplicity, we study particles emitted at $\theta=\pi/2$, i.e,
$p_z=0$
(zero rapidity), and we derive properties of the transverse momentum
distributions. Since the transverse momentum is invariant under
Lorentz boosts along $z$, our final results are valid also at
non-zero rapidity.  

The energy of the particle in the fluid rest frame can be generally
written as $E^*= p^\mu u_\mu$ in the laboratory frame. The reason is
twofold: 1) $p^\mu u_\mu$ is a Lorentz scalar, and is independent of the
frame where it is evaluated; 2) $p^\mu u_\mu$ reduces to $p^0$ if the
fluid velocity is zero. 
Assuming that the fluid velocity is parallel to the particle velocity,
and that $p_z=0$, we obtain
\begin{equation}
E^*=p^\mu u_\mu=m_t u^0-p_t u,
\label{Estar}
\end{equation}
where $u^0=\sqrt{1+u^2}$, 
$p_t$ is the transverse momentum of the particle, and 
$m_t=\sqrt{p_t+m^2}$ its ``transverse mass'', which equals the energy
for a particle with $p_z=0$. 
The definition of a fast particle is that its velocity exceeds the
maximum fluid velocity, i.e., $p_t>m u$ (or equivalently, $m_t>m u^0$)
everywhere. For a fast particle, $E^*$ is minimum if $u$ is maximum:
fast particles are emitted from the regions where the fluid velocity
is largest. 

\subsection{Radial flow}
\label{s:radial}

\begin{figure}
\begin{center}
\includegraphics*[width=0.7\linewidth]{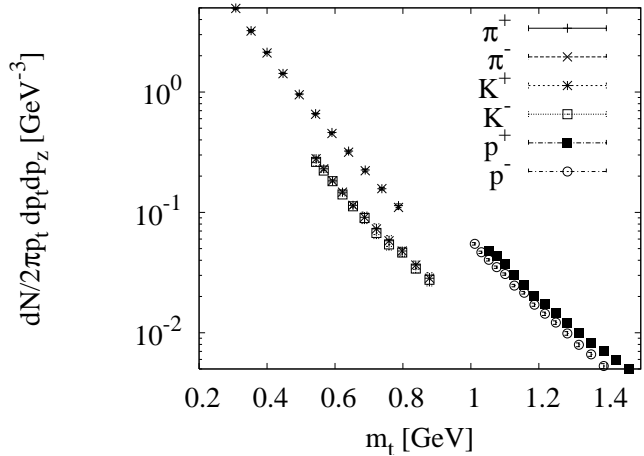}
\end{center}
\caption{$m_t$ spectra of identified hadrons produced in
  p-p collisions near $p_z=0$ 
  (data from \cite{Adams:2003xp}, replotted). 
\label{fig:ppspectra}}
\end{figure}

We first study the transverse momentum distribution of particles
emitted in central collisions. Rotational symmetry in the transverse
plane allows us to write $dp_x dp_y=2\pi p_t dp_t$. 
(\ref{cooperfrye}) and (\ref{Estar}) then give
\begin{equation}
\frac{dN}{2\pi p_t dp_t dp_z}\propto\exp\left(\frac{-m_t u_0+p_t
  u}{T}\right),
\label{mtspectra}
\end{equation}
where $u$ is the maximum transverse fluid 4-velocity at zero rapidity,
according to the above discussion. 
If the fluid is at rest, i.e., $u=0$ and $u^0=1$, one expects that the
spectra are exponential in $m_t$, with the same slope $1/T$ for all
particles. 
It is a general feature of Boltzmann statistics that kinetic energies
associated with thermal motion are always of order $T$, and
independent of the particle mass.
This is precisely what is seen in proton-proton collisions:
figure \ref{fig:ppspectra} displays the momentum
distributions of various hadrons in log scale, as a function of the
transverse mass. $N$ denotes the number of particles per event. 
Pions, kaons, protons and antiprotons are on parallel lines. 
Protons are slightly above antiprotons: this shows that 
the net baryon number is not strictly zero, and that our
``baryonless'' picture is only an approximation. 
The lines of protons and antiprotons are above the line of pions (if
one extrapolates the latter to larger $m_t$),
roughly by a factor 2. This factor 2 corresponds to the spin degrees
of freedom in (\ref{cooperfrye}): $S=\frac{1}{2}$ for protons and 0 for
pions and kaons. 
By contrast, the kaon line is lower than the pion line. This
phenomenon is known as ``strangeness suppression'':
less strange particles are produced in elementary particles 
than expected on the basis of statistical models. 

The fact that thermal models give a satisfactory description of 
particle spectra and abundances in $p-p$
~\cite{Becattini:1997rv}, and even $e^+-e^-$
collisions~\cite{Becattini:1995if} does not prove that thermal
equilibrium is achieved in these collisions.
In fact, thermal equilibrium is not at all expected in such small
systems, and the apparent thermal behaviour still remains a puzzle. 
It could arise from the mechanism of hadronization itself, which is
essentially a statistical process.

\begin{figure}
\begin{center}
\includegraphics*[width=0.7\linewidth]{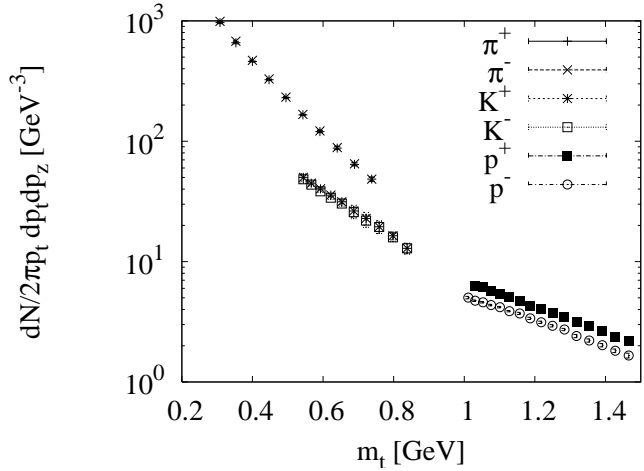}
\end{center}
\caption{$m_t$ spectra of identified hadrons produced in
central  Au-Au collisions near $p_z=0$ 
  (data from \cite{Adams:2003xp}, replotted). 
Yields are normalized per event, which explains why they
are $\approx 200\times$ larger than in p-p collisions. 
\label{fig:Auspectra}}
\end{figure}

We now show that $m_t$-scaling is broken if the fluid moves: on top of
thermal motion, there is now a collective velocity $v$, the fluid
velocity, which applies to all particles within the fluid.  The
kinetic energy associated with this collective motion is $mv^2/2$ in the
nonrelativistic limit. It increases with the particle mass, and one
expects that heavier particles will have larger kinetic energies if
collective flow is present. 
To see the breaking of $m_t$ scaling explicitly, we compute the slope
of the $m_t$ spectrum by taking the log of (\ref{mtspectra}) and
differentiating with respect to $m_t$. We use the fact that
$p_t^2=m_t^2-m^2$ implies $dp_t/dm_t=m_t/p_t$: 
\begin{equation}
\frac{d}{dm_t}\log\left(\frac{dN}{2\pi p_t dp_t
  dp_z}\right)=\frac{-u_0+u m_t/p_t}{T}.
\label{slope}
\end{equation}
For a given $m_t$, 
heavier particles have a smaller $p_t$. If $u>0$, this gives a
positive contribution to the slope, resulting in 
flatter $m_t$-spectra\footnote{Please note that (\ref{slope})
  applies only to fast particles, for which $p_t>m u$ and $m_t>m u_0$,
  so that the slope is always negative.}. This is clearly seen in Au-Au collisions,
figure \ref{fig:Auspectra}: (anti)proton spectra and kaon spectra are
much flatter than pion spectra. This is generally considered
evidence for transverse flow~\cite{Lee:1990sk}. In the case of central
collisions, which have rotational symmetry in the $(x,y)$ plane,
transverse flow is also called ``radial'' flow. 

\subsection{Chemical versus kinetic freeze-out}
\label{s:freezeouts}

Comparing Figs.~\ref{fig:ppspectra} and \ref{fig:Auspectra}, it is
clear that the relative abundances of pions, kaons, and (anti)protons,
also known as particle ratios, do not change dramatically from pp to
Au-Au collisions: what happens between pp and Au-Au is essentially a
redistribution of the transverse masses for heavier particles. 

Now, the number of particles of a given type emitted by a fluid element is
obtained by integrating the Boltzmann factor, (\ref{cooperfrye}),
over momentum. As a consequence, particle ratios only depend on the
temperature. The fact that particle ratios are the same in pp and
Au-Au collisions means that the temperature is the same: the
temperature extracted from particle ratios is called the ``chemical
freeze-out temperature'', and its value is 
$T_c\simeq 170~MeV$~\cite{BraunMunzinger:2001ip}. 
A detailed calculation shows that the kaon/pion ratio is in fact
larger in Au-Au collisions than in pp collisions, and that there is no
``strangeness suppression'' in Au-Au collisions. 
Although this is generally considered a strong argument for
thermalization, it has been shown that the mechanism of quark
production itself can produce apparent thermal
equilibrium~\cite{Gelis:2005pb}.

While the same value of the temperature explains both the particles
ratios and the $m_t$ spectra in pp collisions, it is no longer the
case for Au-Au collisions. If $T$ in (\ref{slope}) was the
same for pp and Au-Au collisions, transverse flow would result in
much flatter pion spectra for Au-Au than pp collisions. The phenomenon
of transverse (or radial) flow nicely explains the slopes of $m_t$
spectra of identified hadrons, but the price to pay is a lower value
of the temperature. 
This temperature is referred to as the temperature of ``kinetic
freeze-out'', and its typical value at RHIC is $T_f\simeq 100$~MeV. 

The fact that $T_f<T_c$ is usually interpreted in the following way:
inelastic collisions, which maintain chemical equilibrium, stop below
$T_c$; below $T_c$, particle abundances are frozen, but there are
still enough elastic collisions to maintain Boltzmann distributions of
momenta, i.e., kinetic equilibrium. Kinetic equilibrium is eventually
broken when the  temperature becomes lower than $T_f$, the kinetic
freeze-out temperature.

\subsection{Elliptic flow}
\label{s:elliptic}

\begin{figure}
\begin{center}
\includegraphics*[width=0.8\linewidth]{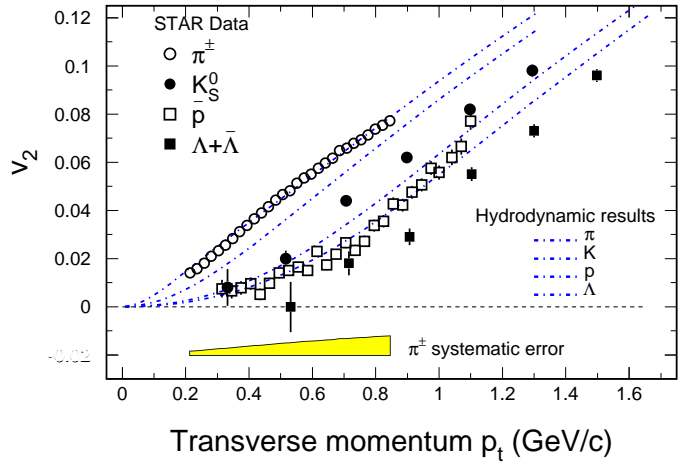}
\end{center}
\caption{Elliptic flow of identified hadrons as a function of
  transverse momentum~\cite{Adams:2004bi}. $v_2$ versus $p_t$ is often
  called ``differential'' elliptic flow. 
\label{fig:v2pt}}
\end{figure}

We now study non-central collisions, and we define the $x$ and $y$
axes as in figure \ref{fig:xy}. We rewrite (\ref{cooperfrye})
using $dp_xdp_y=p_t dp_t d\phi$ and  (\ref{Estar}), where we take
into account the fact that the maximum fluid velocity at zero rapidity
may also depend on $\phi$: 
\begin{equation}
\frac{dN}{p_t dp_t dp_zd\phi}\propto\exp\left(\frac{-m_t u_0(\phi)+p_t
  u(\phi)}{T}\right).
\label{phispectra}
\end{equation}
According to (\ref{vxy}), the fluid velocity is larger on the
$x$-axis than on the $y$-axis, which is the phenomenon referred to as
elliptic flow. This effect can be parameterized in the form
\begin{equation}
u(\phi)=u+ 2 \alpha\cos 2\phi, 
\label{uphi}
\end{equation}
where $\alpha$ is a positive coefficient characterizing the magnitude of
elliptic flow, and $u$ is the average over $\phi$ of the maximum fluid 
4-velocity in the $\phi$ direction.  
In semi-central Au-Au collisions at RHIC, experimental data 
suggest that $\alpha\simeq 4$\%, which means that elliptic flow at the
level of the fluid is a small effect. 
Using $u^0=\sqrt{u^2+1}$, and
expanding to first order in $\alpha$, we obtain 
\begin{equation}
u^0(\phi)=u^0+ 2v\alpha\cos 2\phi, 
\label{u0phi}
\end{equation}
where $v\equiv u/u_0$ is the average over $\phi$ of the maximum fluid
velocity.  
We then insert (\ref{uphi}) and (\ref{u0phi}) into
(\ref{phispectra}) and expand to first order in $\alpha$. Comparing
with (\ref{defv2}), we obtain the value of elliptic flow, $v_2$:
\begin{equation}
v_2= \frac{\alpha}{T}\left(p_t-v m_t\right).
\label{v2pt}
\end{equation}
This equation~\cite{Huovinen:2001cy} explains the essential features
of the differential 
elliptic flow of identified particles, shown in 
figure \ref{fig:v2pt}. For light particles such as pions, $m_t\simeq
p_t$, and $v_2$ increases essentially linearly with $p_t$. This is 
already a non-trivial result. For heavier particles, $m_t$ is larger at
the same value of $p_t$, resulting in smaller $v_2$~\footnote{Please
  note that (\ref{v2pt}) only applies to fast particles, $p_t> mu$
  and $m_t>mu_0$, which implies $v_2>0$.}. This strong mass
ordering is clearly seen in the data: kaons and protons have smaller
$v_2$ than pions at the same $p_t$. (\ref{v2pt}) shows that the 
mass ordering is significant only if $v$ is a significant fraction of
the velocity of light. RHIC data on $v_2$ can therefore be considered
strong evidence for {\it relativistic\/} collective flow. Fits to data
suggest that the maximum fluid velocity may be as large as 0.7.

The increase of $v_2$ with $p_t$ predicted by (\ref{v2pt}) is 
seen in the data only up to $p_t\sim$~2~GeV/c. 
For higher values of $p_t$, $v_2$ saturates and eventually
decreases~\cite{Adams:2004wz}. Such a deviation from ideal
hydrodynamics has been shown to occur generally as a result of viscous
effects~\cite{Teaney:2003kp}, to be discussed briefly in
Sec~\ref{s:last}. However, viscosity alone cannot explain the
observation~\cite{Adler:2003kt} that $v_2$ becomes {\it higher\/} 
for baryons than for mesons between 2 and 3 GeV/c, an effect which 
has been attributed to the process of hadronization through quark 
coalescence~\cite{Molnar:2003ff}.  
This picture has in turn suggested new analyses involving new scaling
variables~\cite{Adare:2006ti}. This illustrates the vivid 
interplay between experiment and phenomenology in the field of
heavy-ion collisions.

\section{Viscosity and thermalization}
\label{s:last}

\subsection{Types of flows}

The various types of flows occurring in fluid mechanics are classified
according to the values of three dimensionless parameters:
\begin{itemize}
\item The Knudsen number $Kn\equiv \lambda/R$ is the ratio of the mean
  free path $\lambda$ of a particle between two collisions, to a
  characteristic spatial dimension of the system, $R$. Applicability of
  hydrodynamics requires $Kn\ll 1$. 
\item The Mach number $Ma\equiv v/c_s$ is the ratio of the
  characteristic flow velocity, $v$, to the sound velocity, $c_s$. It
  can be shown   (see problem 2 in appendix) that if $Ma\ll 1$, the
  density is   almost uniform   throughout the fluid, which defines
  {\it   incompressible\/} flow: whether a fluid is compressible or
  not  depends on how fast it is flowing. 
\item The Reynolds number is defined by $Re\equiv R v/(\eta/\rho)$,
  where $\eta$ is the shear viscosity, and $\rho$ the mass density
  (which must be replaced with $\epsilon+P$ for a relativistic fluid),
  and $R$ and $v$ are defined as above. If $Re\gg 1$, the flow can be
  considered inviscid. 
\end{itemize}
There is a fundamental relation between these three numbers. Transport
theory indeed shows that $\eta/\rho\sim \lambda c_s$, which implies
\begin{equation}
Re\times Kn\sim Ma.
\label{RKM}
\end{equation}
This is a very general relation\footnote{There is a dimensionless
  proportionality constant of order 1 between 
the two sides of (\ref{RKM}), whose precise values depends on the
  interaction. It is $\simeq$~1.6 for
a dilute gas of nonrelativistic hard spheres.}. Since the validity of
  a fluid description requires $Kn\ll 1$, 
(\ref{RKM}) shows that there  are essentially three types of
  flows, which correspond to three  different branches of fluid
  dynamics.  
\begin{itemize}
\item{Compressible flows, for which $Ma$ is of order unity. Since
  $Kn\ll 1$, this in turn implies $Re\gg 1$: compressible flows are
  inviscid. This part of fluid mechanics is called gas dynamics.}
\item{Viscous flows, for which $Re$ is of order unity. Since $Kn\ll
  1$, this in turn implies $Ma\ll 1$: viscous flows are
  incompressible. }
\item{Incompressible, inviscid flows (sometimes called ``ideal''), for
  which $Ma\ll 1$ and $Re\gg 1$. This is where turbulence
 occurs.} 
\end{itemize}
In the case of a heavy-ion collision, the fluid is expanding into the
vacuum: this is obviously a compressible flow, where $Ma$ is of order
unity. The real question is the validity of the fluid description,
i.e., the actual value of $Kn$. 

\subsection{Viscous corrections}

The dynamics of gases expanding into the vacuum has been extensively 
studied in nonrelativistic gas dynamics~\cite{Cercignani}. The Knudsen
number $Kn$ provides a natural small parameter for these problems, and
observables can be computed by an expansion in powers of $Kn$:
\begin{itemize}
\item{The lowest order, i.e., the limit $Kn\to 0$, corresponds to
  inviscid hydrodynamics. }
\item{The first correction, linear in $Kn$, is also linear in the
  viscosity, since $Kn\propto 1/Re\propto\eta$. The corresponding
  fluid equations are Navier-Stokes equations, or viscous
  hydrodynamics. They involve several transport coefficients 
  (diffusion, shear and bulk viscosities), and the energy-momentum
  tensor is no longer symmetric.}
\item{The next correction, in $Kn^2$, is described by more complicated
  equations called the Burnett equations~\cite{burnett,bgk}. }
\end{itemize}
A heavy-ion collision at RHIC produces a few thousand particles. It is
  intuitively obvious that the fluid picture is at best an
  approximation, and that there are sizeable corrections to this 
  picture. The question of whether or not hydrodynamics applies to
  heavy-ion collisions is no longer a qualitative question, but
  rather a quantitative one. 
  This is what viscosity is about: the goal of viscous
  hydrodynamics is to provide a more accurate description of heavy-ion
  collision, by taking into account the leading corrections to the
  ideal-fluid
  picture~\cite{Baier:2006um,Bhalerao:2007ek}. 

We conclude with estimates of the Knudsen number at RHIC. 
The actual value of the viscosity of hot QCD is not known at
present. Estimates have been obtained from lattice
QCD~\cite{Meyer:2007ic} but there are still
controversial. Interestingly, a universal lower bound on the viscosity
to entropy ratio, which might hold for all field theories,  has been
proposed on the basis of a correspondence with black-hole 
physics~\cite{Kovtun:2004de}. This universal bound is
\begin{equation}
\frac{\eta}{s}>\frac{\hbar}{4\pi}. 
\label{KSS}
\end{equation}
This lower bound on $\eta$ can be converted into an upper bound on the
Reynolds number. Since $Kn\sim 1/Re$, this in turn gives a lower bound on the
Knudsen number, which is of order $0.1$ for central Au-Au
collisions. This means that viscous corrections at RHIC are expected
to be 10\% at least. 
A recent study of elliptic flow~\cite{Drescher:2007cd} suggests that
the magnitude of viscous corrections is at least 30\%. 
This in turn would mean that the viscosity of hot QCD is significantly
larger than the KSS bound, (\ref{KSS}).

Inviscid hydrodynamics 
gives a satisfactory explanation of several RHIC data at the qualitative
level: mass ordering of $m_t$ spectra, differential elliptic flow. 
However, they are unable to reproduce all the data quantitatively. 
Taking into account viscous corrections will be a major step in
this respect. This is an ongoing programme. A lot
of progress has already been made, and quantitative results, 
with comparison to RHIC data, are now appearing~\cite{Romatschke:2007mq}. 
Eventually, one should be able to estimate both the equation of
state and the viscosity of hot QCD from heavy-ion experiments. 
Hydrodynamic calculations may even shed light on the initial density
profile, i.e., on the early stages of the collision, and the particle
production itself. Hydrodynamics was crucial in our understanding of
heavy-ion collisions at RHIC. It will be even more important at LHC,
where the quark-gluon plasma will last longer than at RHIC, and the
whole expansion will be dominated by hydrodynamics. 

\ack
I thank IIT Mumbai and TIFR for their hospitality, 
CEFIPRA for financial support under project 3104-3, and F. Grassi and
R. Bhalerao for discussions and useful comments on the manuscript. 

\appendix
\section{Problems}
\subsection*{Problem 1: Equations of inviscid hydrodynamics}

1. We introduce the notations $D\equiv u^\mu\partial_\mu$ and 
$\Delta^{\mu\nu}\equiv g^{\mu\nu}-u^\mu u^\nu$. 
How do these quantities simplify for a fluid at rest?

\noindent
2. Using (\ref{umuumu}), show that $u_\nu\partial_\mu u^\nu=0$ and
   $\Delta_{\mu\nu} u^\nu=0$. 

\noindent
3. Multiply the equations of energy-momentum conservation
   (\ref{tmunu}-\ref{dmutmunu}) by $u_\nu$ 
   and show that $u^\mu\partial_\mu\epsilon+(\epsilon+P)\partial_\mu
   u^\mu=0$. 

\noindent
4. Using (\ref{Extensive3}), (\ref{depsilon}) and
   (\ref{nconservation}), show that $\partial_\mu(s u^\mu)=0$.  
   What is the interpretation of this equation?

\noindent
5. Show that $D(s/n)=0$. What is the interpretation of this result?

\noindent
6. Multiply the equations of energy-momentum conservation by
   $\Delta_{\rho\nu}$ and show that 
\begin{equation}
(\epsilon+P)Du_\rho=\Delta_{\rho\nu}\partial^\nu P. 
\label{eulereq}
\end{equation}
What is the non-relativistic limit of this equation?

\noindent
7. Explain why the previous equation, together with the equation of
   entropy conservation and baryon number conservation, exhausts all
   the information contained in the equations of hydrodynamics. 

\subsection*{Problem 2: Steady flows}
\label{s:steady}

The flow in a heavy-ion collision is strongly time dependent. 
Studying steady flows is somewhat academic in this context. 
However, simple exact results can be easily obtained for steady flows,
and they provide useful insight into the physics of hydrodynamics.

\noindent
1. Show that (\ref{eulereq}) for $\rho=0$, 
in the case of a steady flow (where all quantities are time
independent), can be recast in the form 
\begin{equation}
d\ln u^0= -\frac{dP}{\epsilon+P},
\end{equation}
where the differential is taken along a streamline (i.e., a line
parallel to the fluid velocity). 

\noindent
2. Take the nonrelativistic limit of this result for an incompressible
fluid and comment on the result. 

\noindent
3. For a baryonless fluid, show that $u^0 T$ is constant along a
   streamline. 

\noindent
4. The velocity of sound $c_s$ is defined as
$c_s=\sqrt{dP/d\epsilon}$. Show that the result of Q1 can be rewritten as
\begin{equation}
\frac{du}{u}=-\frac{c_s^2}{v^2} \frac{ds}{s}, 
\end{equation} 
where $v=u/u^0$ is the fluid velocity. The Mach number of a flow is
defined by Ma$\equiv v/c_s$. If Ma$\ll 1$, one says that the flow is 
incompressible. Explain why. 

\noindent
5. Consider an elementary flux tube, and denote by $\Sigma$ the
   cross-section area of the flux tube at some point. Explain why 
$s u\Sigma$ is a constant along the flux tube. Write this in
   differential form.

\noindent
6. Eliminate the fluid 4-velocity $u$ between the results of Q4 and Q5
   and show that
\begin{equation}
\frac{d\Sigma}{\Sigma}=\left(\frac{c_s^2}{v^2}-1\right)\frac{ds}{s}
\end{equation}
along the flux tube. How does the density evolve diverging
streamlines, depending on whether the flow is supersonic or subsonic?

\noindent
7. Consider the case where a nozzle emits a baryonless gas, which then
   expands into the vacuum. List some consequences of the results
   obtained in this problem. 

\subsection*{Problem 3: The Riemann problem}

The Riemann problem is a one-dimensional time-dependent flow which can be solved exactly.
The initial conditions are: at time $t=0$, the half space $x<0$ is filled with a uniform
fluid at rest, with energy density $\epsilon_0$, while the half space $x>0$ is empty.
We shall determine the flow profile at $t>0$. Since there is no characteristic  length or time 
scale in the problem, both the fluid velocity and the density depend through $x$ and $t$ 
only through the combination $\zeta=x/t$: the flow profile has the same shape at all 
positive times, only its size increases linearly with time.

\noindent
1. Sketch the density profile at positive time. 

\noindent
2. We first determine the point where the matter starts to flow to the right. 
At this point the fluid velocity is 0 by continuity, but the 
derivatives of $v$ with respect to $x$ and $t$ are generally not 0. 
(\ref{firstordereqs}) simplify to:
\begin{eqnarray}
\frac{\partial\epsilon}{\partial t}+(\epsilon+P)\frac{\partial
  v}{\partial x}&=&0\cr 
\frac{\partial P}{\partial x}+(\epsilon+P)\frac{\partial v}{\partial
  t}&=&0.
\end{eqnarray}
Rewrive these partial differential equations as ordinary differential equations in the 
reduced variable $\zeta=x/t$.

\noindent
3. Eliminate the pressure from this equation using $dP=c_s^2 d\epsilon$. Show that the 
resulting system of equations has a nontrivial solution only if $\zeta=\pm c_s$. 
In the situation considered here, one expects $d\epsilon/d\zeta<0$ and $dv/d\zeta>0$. 
Show that this implies $\zeta=-c_s$. 
At which value of $\zeta$ does the matter start to flow? 
Comment on this result.

\noindent
4. Since the equations are Lorentz-invariant, at every point one can perform a Lorentz 
boost such that the fluid velocity is 0 in the new frame. Explain, without algebra, why
the above result generalizes to $\zeta=(v-c_s)/(1-vc_s)$ at a point where the fluid velocity
is not zero.

\noindent
5. Invert this relation and draw the velocity profile as a function of $x$ for an 
ideal quark-gluon plasma with sound velocity $c_s=1/\sqrt{3}$.

\section{Solutions}

\subsection{Solution of problem 1}

\noindent
1. $D=u^0(\partial_t+\vec v\cdot\vec\nabla)$ where $\vec v=\vec u/u^0$
   is the fluid velocity. In the nonrelativistic
   limit, $u^0=1$ and $D$ is the convective derivative, i.e.,
   the time derivative along a comoving fluid element. For a fluid at
   rest, $D$ is the time derivative and $\Delta^{\mu\nu}={\rm
   diag}(0,-1,-1,-1)$ projects spacetime onto space. 

\noindent
2. By taking the derivative of $u^\nu u_\nu=1$, one obtains 
$u_\nu\partial_\mu u^\nu=0$. From the definition of $\Delta_{\mu\nu}$
it is obvious that $\Delta_{\mu\nu}u^\nu=0$. 

\noindent
3. The equation of energy-momentum conservation can be written as the
sum of 3 terms:
\begin{equation}
(\epsilon+P)u^\mu\partial_\mu u^\nu+\partial_\mu ((\epsilon+P)u^\mu)
u^\nu-\partial^\nu P=0.
\end{equation}
Multiplying this equation by $u_\nu$, the first term disappears using
the result of Q2. One obtains
\begin{equation}
\partial_\mu((\epsilon+P)u^\mu)-u^\mu\partial_\mu P=0. 
\end{equation}
Expanding the first term, one obtains 
\begin{equation}
u^\mu\partial_\mu\epsilon+(\epsilon+P)\partial_\mu u^\mu=0.
\end{equation}

\noindent
4. (\ref{nconservation}) gives 
\begin{equation}
u^\mu\partial_\mu n+n\partial_\mu u^\mu=0.
\label{n2}
\end{equation}
Multiplying by $\mu$ and subtracting from the previous equation, one
obtains 
\begin{equation}
u^\mu T\partial_\mu s+T s\partial_\mu u^\mu=0. 
\end{equation}
Simplifying by $T$, this can be recast in the form
$\partial_\mu(su^\mu)=0$. This equation is formally analogous to the
equation of baryon number conservation (\ref{nconservation}), where
the baryon number is replaced with the entropy: it expresses entropy
conservation. 

\noindent
5. (\ref{n2}) gives $D n/n=-\partial_\mu u^\mu$. Similarly, the
   equation of entropy conservation gives $D s/s=-\partial_\mu
   u^\mu$. It follows that $D(s/n)=(s/n)(Ds/s-Dn/n)=0$. This equation
   means that the entropy per baryon,  $s/n$, is constant along a
   comoving fluid element. 

\noindent
6. Again, write the equation of energy-momentum conservation as the 
sum of 3 terms:
\begin{equation}
(\epsilon+P)u^\mu\partial_\mu u^\nu+\partial_\mu ((\epsilon+P)u^\mu)
u^\nu-\partial^\nu P=0.
\end{equation}
Multiply by $\Delta_{\rho\nu}$, the second term disappears
and $\Delta_{\rho\nu}\partial_\mu u^\nu=\partial_\mu u_\rho$. 
One thus obtains immediately
\begin{equation}
(\epsilon+P)Du_\rho=\Delta_{\rho\nu}\partial^\nu P. 
\end{equation}
In the nonrelativistic limit, $\Delta_{\rho\nu}$ projects onto the
space components, so that $\Delta_{\rho\nu}\partial^\nu P$ is the
pressure gradient. 
$\epsilon+P$ reduces to the mass density and one recovers Euler's
equation, i.e., Newton's second law of motion applied to the fluid
element. 

\noindent
7. In Q3 we have projected the equations on the time-like direction
   $u^\mu$, in Q6 we have projected on space. All the information has
   been used.

\subsection{Solution of problem 2}

\noindent
1. The equation for $\rho=0$ is 
\begin{equation}
(\epsilon+P)Du_0=\Delta_{0\nu}\partial^\nu P.
\end{equation}
For a stationary flow, $\partial^0 P=0$, and only the spatial
components remain on the right-hand side. 
$D$ reduces to $\vec u\cdot\nabla$. 
Inserting the definition of
$\Delta$, one obtains
\begin{equation}
(\epsilon+P)\vec u\cdot\vec\nabla u_0=-u_0 u_i\partial^i P=-u_0 \vec
u\cdot\vec\nabla P. 
\end{equation}
Dividing both sides by $u_0$, and writing 
$\vec u\cdot\vec\nabla=u (d/d\sigma)$, where $d\sigma$ is the length along a
steamline, one obtains the result.

\noindent
2. In the nonrelativistic limit, $u_0\simeq 1+\vec v^2/(2c^2)$: to
   leading order in $\vec v$, $\ln u_0=\vec v^2/(2 c^2)$. Next,
   $\epsilon+P=\rho c^2$, where $\rho$ is the mass density. For an
   incompressible fluid, this shows that $v^2/2+P/\rho$ is a constant
   along a streamline. This is Bernoulli's equation, which states that
   when the fluid accelerates, the pressure decreases. This equation
   has many  applications in fluid dynamics; it explains how a
   tornado can lift objects. 

\noindent
3. For a baryonless fluid, (\ref{Extensive3}) and (\ref{dP}) give
   $dP/(\epsilon+P)=dT/T=d\ln T$. The relativistic
   Bernoulli equation then becomes $d\ln u^0+d\ln T=0$ along a
   streamline, from which one easily proves the result. The fluid
   cools as it accelerates. 

\noindent
4. $u_0^2-u^2=1$, hence $u_0 du_0=udu$. This implies
   $du/u=(du_0/u_0)/v^2$. We then write $dP=c_s^2 d\epsilon$
   (\ref{defcs}), and $d\epsilon/(\epsilon+P)=ds/s$
   (\ref{isentropic}). This gives the   result.  

\noindent
5. Conservation of entropy implies that the entropy flux is constant
along the flux tube. This implies that $s u\Sigma$ is constant.
In differential form, this writes
\begin{equation}
\frac{ds}{s}+\frac{du}{u}+\frac{d\Sigma}{\Sigma}=0.
\end{equation}

\noindent
6. Replacing $du/u$ with the result of Q4 in the above equation gives
   the result. If the streamlines diverge, $d\Sigma/\Sigma$ is
   positive. For a supersonic flow, $v>c_s$, this implies $ds>0$,
   i.e., the density decreases along the streamline. For a subsonic
   flow, it increases.

\noindent
7. For a gas expanding into the vacuum, streamlines obviously diverge,
   and the density decreases. This means that the flow is
   supersonic. As the fluid cools, it accelerates, $u_0\propto
   1/T$. As the fluid becomes cooler and cooler, 
the mean free path becomes eventually too large for hydro to
   be valid. This occurs when the flow is ultrarelativistic, i.e.,
   $u^0\gg 1$. In the nonrelativistic case, this condition becomes
   $v\gg c_s$, and this is called ``hypersonic flow''. 

\subsection{Solution of problem 3}

\noindent
1. One expects the matter to flow to the right, so that the density will smoothly 
decrease as a function of $x$. Since the information cannot propagate faster than the 
speed of light, one expects that the density is $\epsilon_0$ for $x<-t$, and 0 for
$x>t$. The flow occurs in the interval $-t<x<t$.

\noindent
2. One simply does the replacements $\partial/\partial x=(1/t)d/d\zeta $, 
$\partial/\partial t=-(\zeta /t)d/d\zeta $. The system of equations becomes
\begin{eqnarray}
-\zeta \frac{d\epsilon}{d \zeta }+(\epsilon+P)\frac{d v}{d \zeta }&=&0\cr
\frac{d P}{d \zeta }-\zeta (\epsilon+P)\frac{d v}{d \zeta }&=&0.
\end{eqnarray}

\noindent
3. Replacing $dP$ with $c_s^2 d\epsilon$, one obtains a linear system of 2 equations
with unknowns $d\epsilon/d\zeta $ and $dv/d\zeta $. The system has a trivial solution
$dv/d\zeta =d\epsilon/d\zeta =0$. It has nontrivial solutions only if the determinant vanishes, 
which gives $\zeta ^2=c_s^2$, i.e. $\zeta =\pm c_s$. The conditions  
$d\epsilon/d\zeta <0$ and $dv/d\zeta >0$ imply $\zeta <0$ (see equations above). The correct solution
is therefore $\zeta =-c_s$. The matter starts to flow at $x=-c_s t$. 
For $x<-c_s t$, the flow velocity is 0 and the density is equal to the initial value
$\epsilon_0$, corresponding to the trivial solutions of the hydrodynamic equations. 
 
\noindent
4. In the frame where the fluid velocity is zero, $\zeta =-c_s$, which
means that the information travels at velocity $-c_s$ with respect to
the fluid. Under a Lorentz boost of velocity $v$, the relativistic addition of 
velocities applies, so that $\zeta =(v-c_s)/(1-vc_s)$.

\begin{figure}
\begin{center}
\includegraphics*[width=0.7\linewidth]{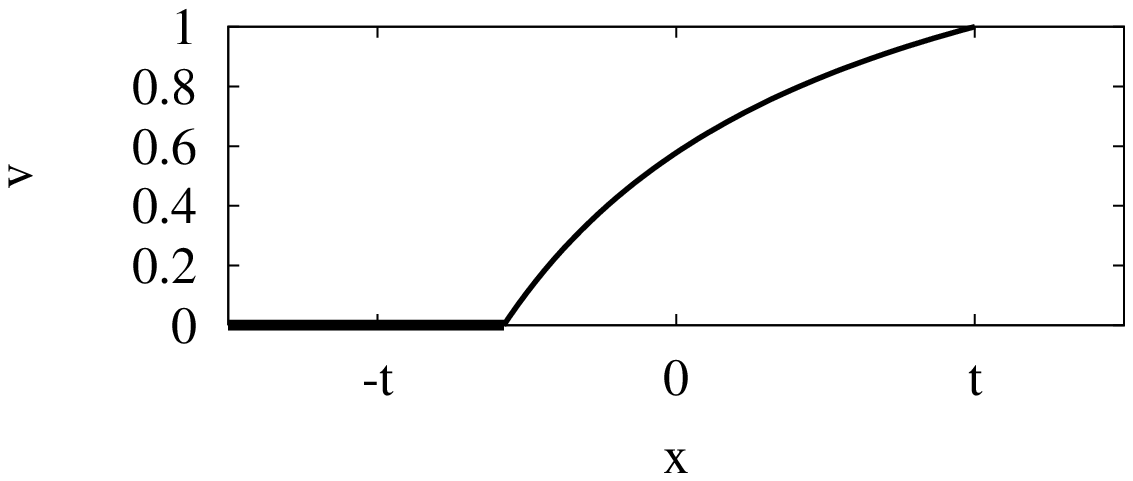}
\end{center}
\caption{Velocity profile for the Riemann problem. 
\label{fig:riemann}}
\end{figure}

\noindent
5. Inverting the relation, we obtain $v=(\zeta +c_s)/(1+\zeta c_s)$. The maximum value of $v$ is 1,
which corresponds to $\zeta =1$. Note that the fluid velocity at $x=0$
is exactly $c_s$. The velocity profile is shown in figure \ref{fig:riemann}.

\end{document}